\documentclass[prd,twocolumn,superscriptaddress,showpacs,nofootinbib,preprintnumbers]{revtex4}

\usepackage{amsmath}
\usepackage{amsfonts}
\usepackage{graphicx}
\usepackage{dcolumn}
\usepackage{hyperref}
\newcommand{\be}{\begin{equation}}
\newcommand{\ee}{\end{equation}}
\newcommand{\bea}{\begin{eqnarray}}
\newcommand{\eea}{\end{eqnarray}}
\newcommand{\beaa}{\begin{eqnarray*}}
\newcommand{\eeaa}{\end{eqnarray*}}

\newcommand{\nn}{\nonumber \\}
\newcommand{\e}{{\rm e}}

\begin{document}

\tolerance=5000

\title{Modified Gauss-Bonnet theory as gravitational alternative for dark
energy}

\author{Shin'ichi Nojiri}
\thanks{Electronic address: snojiri@yukawa.kyoto-u.ac.jp, \\ nojiri@cc.nda.ac.jp},
\affiliation{Department of Applied Physics, National Defence Academy,
Hashirimizu Yokosuka 239-8686, Japan}
\author{Sergei D. Odintsov}
\thanks{Electronic address: odintsov@ieec.uab.es}
\affiliation{Instituci\`{o} Catalana de Recerca i Estudis Avan\c{c}ats
(ICREA) and Institut d' Estudis Espacials de Catalunya (IEEC/ICE),
Edifici Nexus, Gran Capit\`{a} 2-4, 08034 Barcelona, Spain}

\begin{abstract}
We suggest the modified gravity where some arbitrary function of
Gauss-Bonnet (GB)
term is added to Einstein action as gravitational dark energy. It is shown
that such theory may pass
solar system tests. It is demonstrated that modified GB gravity may
describe the most interesting features of late-time cosmology:
the transition from deceleration to acceleration, crossing the phantom
divide, current acceleration  with effective (cosmological
constant, quintessence or phantom) equation of state of the universe.

\end{abstract}

\pacs{98.80.-k, 98.80.Es, 97.60.Bw, 98.70.Dk}

\maketitle

\noindent
1. The explanation of the current acceleration of the universe (dark energy
problem) remains to be a challenge for theoretical physics.
Among the number of the approaches to dark energy, the very interesting
one is related with the modifications of gravity at large distances.
For instance, adding $1/R$ term \cite{salvatore,CDTT} to Einstein
action leads to gravitational alternative for dark energy where late-time
acceleration is caused by the universe expansion. Unfortunately, such
$1/R$ gravity contains
the instabilities \cite{DK} of gravitationally bound objects. These
instabilities may disappear with the account of
higher derivative terms  leading to consistent modified gravity\cite{fR}.
Another proposals for modified gravity
suggest $lnR$ \cite{lnR} or $Tr1/R$ terms\cite{Tr}, account of inverse
powers of
Riemann invariants \cite{carroll} or some other modifications
\cite{other}. The one-loop quantization of general $f(R)$ in de Sitter
space is also done
\cite{quantum}. In addition to the stability condition which significally
restricts the possible form of $f(R)$ gravity, another restriction
comes from the study of its newtonian limit \cite{newton}.
Passing these two solar system tests leads to necessity of fine-tuning
of the form and coefficients in $f(R)$ action, like in consistent modified
gravity \cite{fR}. That is why it has been even suggested to consider such
alternative gravities in Palatini formulation (for recent discussion and
list of references, see \cite{palatini}).

In the present paper we suggest new class of modified gravity,
where Einstein action is modified by the function $f(G)$, $G$ being
Gauss-Bonnet (GB) invariant. It is known that $G$ is topological invariant
in four dimensions while it may lead to number of interesting
cosmological effects
in higher dimensional brane-world approach (for review, see \cite{NOO}).
It naturally appears
in the low energy effective action from string/M-theory (for recent
discussion of late-time cosmology in stringy gravity with GB term, see
\cite{sami}). As we demonstrate below, modified $f(G)$ gravity passes
solar system tests for  reasonable choice of the function $f$.
Moreover, it is shown that such modified GB gravity may describe
late-time (effective quintessence, phantom or cosmological constant)
acceleration of
the
universe. For quite large class of functions $f$ it is possible to
describe
the transition from deceleration to acceleration or from non-phantom phase
to phantom phase in the late universe within such theory.
Thus, modified GB gravity represents quite interesting gravitational
alternative
for dark energy with  more freedom if compare with $f(R)$ gravity.

\noindent
2. Let us start from the following action:
\be
\label{GB1}
S=\int d^4 x\sqrt{-g}\left(\frac{1}{2\kappa^2}R + f(G)\right)\ .
\ee
Here $G$ is the GB invariant:
$G=R^2 -4 R_{\mu\nu} R^{\mu\nu} + R_{\mu\nu\xi\sigma} R^{\mu\nu\xi\sigma}$.
By introducing two auxilliary fields $A$ and $B$, one may rewrite the
action (\ref{GB1}) as
\be
\label{GB3}
S=\int d^4 x\sqrt{-g}\left(\frac{1}{2\kappa^2}R + B\left(G-A\right) + f(A)\right)\ .
\ee
Varying over $B$, it follows $A=G$.
Using it in (\ref{GB3}),  the action  (\ref{GB1}) is recovered.
On the other hand, by the variation over $A$ in (\ref{GB3}), one gets
$B=f'(A)$.
Hence,
\be
\label{GB6}
S=\int d^4 x\sqrt{-g}\left(\frac{1}{2\kappa^2}R + f'(A)G - Af'(A) + f(A)\right)\ .
\ee
The scalar is not dynamical, it has no kinetic term and is introduced for
simplicity.
Varying over $A$, the relation $A=G$ is obtained again.

The spatially-flat FRW universe metric is chosen as
\be
\label{FRW}
ds^2=-dt^2 + a(t)^2 \sum_{i=1}^3  \left(dx^i\right)^2\ .
\ee
The first FRW equation has the following form:
\be
\label{GB7}
0=-\frac{3}{\kappa^2}H^2 + Af'(A) - f(A) - 24 \dot Af''(A) H^3\ .
\ee
Here the Hubble rate $H$ is defined by $H\equiv \dot a/a$.
For (\ref{FRW}), GB invariant $G$ ( $A$) has the
following form:
\be
\label{GB8}
G=A=24\left(\dot H H^2 + H^4\right)\ .
\ee
In general, Eq.(\ref{GB7}) has deSitter universe solution
 where $H$ and therefore $A=G$ are constants.
If  $H=H_0$ with constant $H_0$, Eq.(\ref{GB7}) looks as:
\be
\label{GB7b}
0=-\frac{3}{\kappa^2}H_0^2 + 24H_0^4 f'\left( 24H_0^4 \right) - f\left( 24H_0^4\right) \ .
\ee
For large number of choices of the function $f$, Eq.(\ref{GB7b}) has a
non-trivial
($H_0\neq 0$) real solution for $H_0$ (deSitter universe). Hence, such
deSitter solution may be applied for description of the early-time
inflationary as
well as late-time accelerating universe.

Let us check now how modified GB gravity passes the solar system tests.
The GB correction to the Newton law may be found from the coupling matter
to the action (\ref{GB1}).
Varying over $g_{\mu\nu}$, we obtain
\bea
\label{GB4b}
&& 0= \frac{1}{2\kappa^2}\left(- R^{\mu\nu} + \frac{1}{2} g^{\mu\nu} R\right)
+ T^{\mu\nu} + \frac{1}{2}g^{\mu\nu} f(G) \nn
&& -2 f'(G) R R^{\mu\nu} + 4f'(G)R^\mu_{\ \rho} R^{\nu\rho} \nn
&& -2 f'(G) R^{\mu\rho\sigma\tau}R^\nu_{\ \rho\sigma\tau}
 -4 f'(G) R^{\mu\rho\sigma\nu}R_{\rho\sigma} \nn
&& + 2 \left( \nabla^\mu \nabla^\nu f'(G)\right)R
 - 2 g^{\mu\nu} \left( \nabla^2 f'(G)\right)R \nn
&& - 4 \left( \nabla_\rho \nabla^\mu f'(G)\right)R^{\nu\rho}
 - 4 \left( \nabla_\rho \nabla^\nu f'(G)\right)R^{\mu\rho} \nn
&& + 4 \left( \nabla^2 f'(G) \right)R^{\mu\nu}
+ 4g^{\mu\nu} \left( \nabla_{\rho} \nabla_\sigma f'(G) \right) R^{\rho\sigma} \nn
&& - 4 \left(\nabla_\rho \nabla_\sigma f'(G) \right) R^{\mu\rho\nu\sigma} \ .
\eea
Here $T^{\mu\nu}$ is  matter EMT.
%%%%%%%%%%%
In the expression (\ref{GB4b}), the third derivative $f'''(G)$ is included as
$\nabla^2 f'(G)=f'''(G)\nabla^2 G + f''(G)\nabla^\mu G \nabla_\mu G$, for example.
In the equation (\ref{GB7}) corresponding to the first FRW equation, however, the terms
including $f'''$ do not appear.
%%%%%%%%%%%
When $T^{\mu\nu}=0$, the $(t,t)$-component of Eq.(\ref{GB4b}) reproduces
Eq.(\ref{GB7})
by identifying $G$ with $A$.
The perturbation around the deSitter background which is a solution of
(\ref{GB7b}) may be easily constructed.
We now write  the deSitter space metric as $g_{(0)\mu\nu}$, which gives
the following Riemann tensor:
\be
\label{GB35}
R_{(0)\mu\nu\rho\sigma}=H_0^2\left(g_{(0)\mu\rho}g_{(0)\nu\sigma}
 - g_{(0)\mu\sigma}g_{(0)\nu\rho}\right)\ .
\ee
The flat background corresponds to the limit of $H_0\to 0$.
Represent $g_{\mu\nu}=g_{(0)\mu\nu} + h_{\mu\nu}$.
For simplicity,  the following gauge condition is chosen:
$g_{(0)}^{\mu\nu} h_{\mu\nu}=\nabla_{(0)}^\mu h_{\mu\nu}=0$.
Then Eq.(\ref{GB4b}) gives
\be
\label{GB38}
0=\frac{1}{4\kappa^2} \left( \nabla^2 h_{\mu\nu}
 - 2H_0^2 h_{\mu\nu}\right)
+ T_{\mu\nu}\ .
\ee
The GB term contribution  does not appear except the
length parameter $1/H_0$ of the deSitter space which is determined with
the account of GB term.
 Eq.(\ref{GB38}) proves that there is no correction to Newton law in
deSitter and
even in the flat background corresponding to $H_0\to 0$ whatever is the
form of $f$.
%%%%%%%%%%%%
We should note that the expression (\ref{GB38}) could be valid only in the deSitter background.
In more general FRW universe, there could appear the corrections coming from $f(G)$ term.
We should also note that in deriving (\ref{GB38}), we have used a gauge condition
$g_{(0)}^{\mu\nu} h_{\mu\nu}=0$ but if we include the mode corresponding to
$g_{(0)}^{\mu\nu} h_{\mu\nu}$, there might appear corrections from $f(G)$ term.
Eq.(\ref{GB38}) only shows that for the mode corresponding to the usual
graviton, any
correction coming from $f(G)$ does not appear in the deSitter background.
%%%%%%%%%%%%

In case of $f(R)$-gravity \cite{CDTT,fR}, for most interesting choices of
the function $f$
 an instability was observed in \cite{DK}.
The instability is generated since the FRW equation,
contains derivatives of the fourth order and therefore the scalar
curvature propagates. This may cause the appearence of growing force
between the galaxies.
Only special forms of $f(R)$ may be free of such instability \cite{fR}.

In $f(G)$-gravity case (\ref{GB1}),  the scalar field denoted
by $A$ in (\ref{GB6}) has
no kinetic term. Then the scalar field or the curvature itself does not
propagate and therefore
there is no such instability as in \cite{DK}, which is clear from the (no)
correction to
the Newton law in (\ref{GB4b}). Thus, modified GB gravity may pass
the solar system tests (at least, for some functions $f(G)$).

Having in mind the fundamental property of the current accelerating
universe,
it is interesting to study
 the possible transition between deceleration and acceleration of
the universe
for the action (\ref{GB1}). First, Eq.(\ref{GB8}) could
be rewritten as
$G=A=24H^2 \ddot a/a$.
Then at the transition point $\ddot a=0$, the Gauss-Bonnet term vanishes: $G=A=0$.
Let us assume the transition occurs at $t=t_0$.
The Hubble rate $H$ may be expanded as
\bea
\label{GB40}
H&=&H_0 + H_1\left(t-t_0\right) + H_2\left(t-t_0\right)^2 \nn
&& + H_3\left(t-t_0\right)^3 + {\cal O}\left(\left(t-t_0\right)^4\right)\ .
\eea
At the transition point $t=t_0$, one finds
\be
\label{GB41}
G=48H_0^2 \left( H_0^2 + H_1\right)\ .
\ee
Since $G=0$ there, it follows
\be
\label{GB42}
H_1=- H_0^2\ .
\ee
Hence, $G$ ($A$) is:
\bea
\label{GB43}
G&=&96H_0^2\left(H_2 - H_0^3\right)\left(t-t_0\right) \nn
&& + 48H_0^2\left(5H_0^4 - 2H_0 H_2 + 3H_3\right)\left(t-t_0\right)^2 \nn
&& + {\cal O}\left(\left(t-t_0\right)^3\right)\ .
\eea
We now also assume that $f(A)$ could be expanded as
\be
\label{GB44}
f(A) = f_0 + f_1 A + f_2 A^2 + f_3 A^3 + {\cal O}\left( A^4 \right)\ .
\ee
Then substituting (\ref{GB40}), (\ref{GB43}), and (\ref{GB44}) into
 (\ref{GB7}), we find
\bea
\label{GB45}
H_2&=& H_0^3 - \frac{1}{96H_0^3 f_2}\left(\frac{1}{16\kappa^2} + \frac{f_0}{48 H_0^2}\right)\ ,\nn
H_3&=& - H_0^4 - \frac{1}{96\times 16 \kappa^2 f_2 H_0^2} \nn
&& - \frac{f_3}{96 f_2^3 H_0^4}\left(\frac{1}{16\kappa^2} + \frac{f_0}{48 H_0^2}\right)^2\ .
\eea
Combining (\ref{GB42}) and (\ref{GB45}), one can show that the Hubble rate
can be determined
consistently, which suggests the existence of the transition between
deceleration and acceleration
of the universe.
We should note that $H_0$ could be determined by a proper initial condition.
Then the  transition condition could be  $f_2\neq 0$, that
is, $f(A)$
contains the quadratic term on $A$.

Let us consider now the possible transition between non-phantom phase and
phantom phase of the universe (if current universe is phantom one).
We now assume that the transition occurs at $t=t_1$. Since $\dot H=0$
at the transition point,
it is natural to assume the Hubble rate behaves as
\be
\label{GB46}
H=\tilde H_0 + \tilde H_1 \left(t- t_1\right)^2 + {\cal O}\left(\left(t- t_1\right)^3\right)\ .
\ee
Hence, $A$ ($G$) behaves as
\be
\label{GB47}
G=A=24H_0^4 + 48 H_0^2 H_1 \left(t- t_1\right) + {\cal O}\left(\left(t- t_1\right)^2\right)\ .
\ee
 Eq. (\ref{GB7}) shows
\be
\label{GB48}
\tilde H_1=\frac{- \frac{3}{\kappa^2}\tilde H_0^2 + 24\tilde H_0^4 f'\left(24\tilde H_0^4\right)
 - f\left(24\tilde H_0^4\right)}{1152 \tilde H_0^7 f''\left(24\tilde H_0^4\right)}\ ,
\ee
if $f''\left(24\tilde H_0^4\right)\neq 0$.
We also note that $\tilde H_0$ and $\tilde H_1$ should be positive.
Eq.(\ref{GB48}) suggests the existence of the consistent solution.
Then the crossing of phantom divide
 could occur for large class of functions $f(G)$.
The condition could be $f''\neq 0$ and $\tilde H_1>0$ when $H=\tilde H_0>0$.

The transition between non-phantom phase and phantom phase could be regarded
as a perturbation from the deSitter solution in (\ref{GB7b}).
The following perturbation may be suggested
\be
\label{GB49}
H=H_0 + \delta H\ ,
\ee
where $H_0$ satisfies (\ref{GB7b}).
 Eq. (\ref{GB7}) gives
\bea
\label{GB50}
0&=&\delta \ddot H + 3H_0 \delta \dot H \nn
&& + \left(- 4H_0^2 + \frac{1}{96 \kappa^2 H_0^4 f''\left(24H_0^4\right)}\right) \delta H\ .
\eea
Here, it is supposed that $f''\left(24H_0^4\right)\neq 0$.
Let $\lambda$ satisfies
\be
\label{GB51}
0=\lambda^2 + 3H_0 \lambda - 4H_0^2 +
\frac{1}{96 \kappa^2 H_0^4 f''\left(24H_0^4\right)} \ .
\ee
Then $\delta H$ behaves as $\delta H \sim \e^{\lambda t}$.
If $\lambda$ is real, $\delta H$ is monotonically increasing or
decreasing,
$\dot H$ does not vanish and therefore there is no transition.
%The determinant of (\ref{GB51}) is given by
%\be
%\label{GB52}
%D=25H_0^4\left(1 - \frac{1}{600 \kappa^2 H_0^8
%f''\left(24H_0^4\right)}\right)\ .
%\ee
If
\be
\label{GB53}
600 \kappa^2 H_0^8 f''\left(24H_0^4\right)<1\ ,
\ee
$\lambda$ becomes imaginary and $\delta H$ oscillates.
Therefore the transition between the phantom phase,
where $\dot H=\delta \dot H>0$,
and the non-phantom phase, where $\dot H = \delta \dot H<0$ could be
repeated in oscillation regime.
%Since
%\be
%\label{GB54}
%\lambda = \frac{-3H_0 \pm i \sqrt{-D}}{2}\ ,
%\ee
The amplitude of the oscillation of $\delta H$ decreases as $\left| \delta
H\right|\sim
\e^{-3H_0t/2}$ and the universe asymptotically goes to deSitter space.

\noindent
3. To show that modified GB gravity may lead to quite rich and realistic
cosmological dynamics we consider some explicit examples of the function
$f(G)$.
The following solvable model, which does not belong to the above class,
may be discussed
\be
\label{GB9}
f(A)=f_0\left|A\right|^{1/2}\ .
\ee
Here $f_0$ is a constant.
Assuming
\be
\label{GB10}
H=\left\{
\begin{array}{ll}
\frac{h_0}{t}\quad & \left(\mbox{when}\ h_0>0\right) \\
 -\frac{h_0}{t_0 - t}\quad &\left(\mbox{when}\ h_0<0\right)
\end{array}\right. \ ,
\ee
 Eq.(\ref{GB7}) gives
\be
\label{GB11}
0=-\frac{3}{\kappa^2}h_0^2 - \frac{f_0\left(1+h_0\right)}{2\left(h_0 - 1\right)}
\left|24h_0^3\left(h_0 - 1\right)\right|^{1/2} \ .
\ee
The solution differs for $h_0>1$ and $h_0<0$ cases.
%\bea
%\label{GB12}
%F_1\left(h_0\right) &\equiv& \left(2f_0^2 +
%\frac{3}{\kappa^4}\right)h_0^2
%+ \left(4f_0^2 + \frac{3}{\kappa^4}\right)h_0 + 2 f_0^2 \nn
%&=& 0\ ,
%\eea
%and if we assume $0<h_0<1$,
%\bea
%\label{GB13}
%F_2\left(h_0\right) &\equiv& \left(2f_0^2 +
%\frac{3}{\kappa^4}\right)h_0^2
%+ \left(4f_0^2 - \frac{3}{\kappa^4}\right)h_0 + 2 f_0^2 \nn
%&=& 0\ ,
%\eea
%As the equation $0=F_1\left(h_0\right)$ has always positive
%determinant $D_1$:
%\be
%\label{GB14}
%D_1=\frac{48f_0^2}{\kappa^4} + \frac{9}{\kappa^8}\ ,
%\ee
%there are always two real solutions for $h_0$.
The simple analysis shows that when $f_0^2>3/2\kappa^4$,
there are two negative
solutions, which decribe effective phantom universe with $w<-1$.
When $f_0^2<3/2\kappa^4$, we have one $h_0>1$ solution, which describes
the effective quinteessence $-1<w<-1/3$ and
one $h_0<0$ solution describing effective phantom.
%On the other hand, the determinant $D_2$
%of the equation $F_2\left(h_0\right)=0$ is given by
%\be
%\label{GB15}
%D_2=-\frac{48f_0^2}{\kappa^4} + \frac{9}{\kappa^8}\ .
%\ee
%Then in order that the equation $F_2\left(h_0\right)=0$ has real
%solution, we require
If $16f_0/3\kappa^4 < 1/\kappa^8$,
there are two real solutions which satisfy $0<h_0<1$.

For the second model let $f(A)$ behaves as
\be
\label{GB17}
f(A)\sim f_0 \left| A \right|^\beta\ ,
\ee
in a proper limit with constants $f_0$ and $\beta$.
We now assume in a limit $t\to \infty$, $H$ behaves as
\be
\label{GB18}
H\sim h_0 t^\alpha\ ,
\ee
with constants $h_0$ and $\alpha$. If  $\alpha>-1$, $A$ ($G$) behaves as
\be
\label{GB19}
A=G\sim 24 h_0^4 t^{4\alpha}\ .
\ee
Then (\ref{GB7}) gives $4\alpha\beta=2\alpha$, i.e. $\beta=1/2$ or
$\alpha=0$.
The case $\beta=1/2$ corresponds to the model (\ref{GB9}). On the
other hand, since $H$ is a constant when $\alpha=0$, this case corresponds to (\ref{GB7b}).
Hence,  the only new case is $\alpha<-1$, where
\be
\label{GB21}
A=G\sim \alpha h_0^3 t^{3\alpha -1}\ .
\ee
Then in (\ref{GB7}),  the behavior of each term can be described as
$H^2\sim {\cal O}\left(t^{2\alpha}\right)$,
$Af'(A)\sim f(A) \sim {\cal O}\left(t^{(3\alpha-1)\beta}\right)$,
$\dot Af''(A)H^3 \sim {\cal O}\left(t^{\alpha\left(3\beta + 1\right)}\right)$.
%If we require $2\alpha = (3\alpha-1)\beta = \alpha\left(3\beta +
%1\right)$, we find $\beta=-\alpha$
%and $\alpha=0, -1/3$, which contradicts with the assumption $\alpha<-1$.
%Even if we require $2\alpha < (3\alpha-1)\beta = \alpha\left(3\beta +
%1\right)$
%or $2\alpha = \alpha\left(3\beta + 1\right) > (3\alpha-1)\beta $, we can
%find there is no solution.

With the assumption $2\alpha = (3\alpha-1)\beta > \alpha\left(3\beta +
1\right)$, one finds
$\beta=2\alpha/\left(3\alpha -1\right)$ and $\alpha<-1/3$, which is consistent
with  $\alpha<-1$. The assumption $\alpha<-1$
and $\beta=2\alpha/\left(3\alpha -1\right)$ restricts $1/2<\beta<2/3$.
Eq.(\ref{GB7}) also gives
\be
\label{GB24}
0=-\frac{3h_0^2}{\kappa^2} + \left(\beta - 1\right)f_0\left|\alpha h_0^3\right|^\beta\ ,
\ee
which leads to  non-trivial solution for $h_0$ if $f_0<0$.

Let us consider $t\to 0$ or $t_0 - t\to 0$ limit.
If  $\alpha>-1$, we obtain Eq.(\ref{GB21}).
If one requires $2\alpha = (3\alpha-1)\beta = \alpha\left(3\beta +
1\right)$, then $\beta=-\alpha$
and $\alpha=0, -1/3$. In case $\beta=-\alpha = 1/3$, Eq.(\ref{GB7}) tells
\be
\label{GB25}
0=-\frac{3}{\kappa^2} + \frac{94}{3}f_0 \left|\frac{h_0^3}{3}\right|^{1/3}\ ,
\ee
which requires $f_0>0$ and gives a non-trivial solution for $h_0$.

%If we require $2\alpha > (3\alpha-1)\beta = \alpha\left(3\beta +
%1\right)$, we find $\beta=-\alpha$
%but Eq.(\ref{GB7}) tells $\alpha = 25/72$ or $\alpha=-1$, which
%contradicts
%with the assumption $\alpha>-1$.
%If we assume $2\alpha = (3\alpha-1)\beta < \alpha\left(3\beta +
%1\right)$, we find
%$-1/3<\alpha<0$ or $\alpha>1/3$, which contradicts
%with the assumption $\alpha>-1$ again.

If one restricts $2\alpha = \alpha\left(3\beta + 1\right) <
(3\alpha-1)\beta $,
it follows $\beta=-1/3$. In this case Eq.(\ref{GB7}) gives
\be
\label{GB26}
0=-\frac{3}{\kappa^2}h_0^2 - \frac{32(3\alpha -1)f_0}{3\alpha}\left|\alpha h_0^3\right|^{-1/3}\ .
\ee
where $\alpha$ can be arbitrary as long as $\alpha>-1$.

If $\alpha<-1$ (see Eq.(\ref{GB19})),
Eq.(\ref{GB7}) gives
\bea
\label{GB27}
\beta &=& \frac{1}{2} + \frac{1}{4\alpha}\ ,\nn
0 &=& -\frac{3}{\kappa^2} + 24(4\beta - 5)\beta (\beta -1)
\left|\frac{h_0^3}{2(2\beta -1)}\right|^\beta\ .
\eea
The first equation shows $1/2>\beta>1/4$.
The second equation in (\ref{GB27}) gives a non-trivial solution for $h_0$
if $f_0>0$.

The above results show that the asymptotic solution behaving as
(\ref{GB18}) with $\alpha\neq 0$
can be obtained only for the cases where $\beta=1/2$ or $1/2<\beta<2/3$
when $t\to \infty$
or $\beta=-1/3$ or $1/4<\beta<1/2$ when $t$ or $t_0 -t \to 0$. In other
situations, the asymptotic
solution corresponds to the deSitter space  (\ref{GB7b}).
The following model may be taken as next example:
\be
\label{GB29}
f(A)=-\frac{\alpha}{A} + \beta A^2\ .
\ee
Then Eq.(\ref{GB7b}) gives
\be
\label{GB30}
0=-\frac{3}{\kappa^2}H_0^2 + \frac{\alpha}{12H_0^4} + 576 \beta H_0^8\ .
\ee
It could be difficult to solve the above equation exactly, but
if  the curvature
and therefore $H_0$ is small, one gets
\be
\label{GB31}
H_0^6\sim H_s^6\equiv \frac{\alpha\kappa^2}{3}\ .
\ee
On the other hand if the curvature is large,
we obtain
\be
\label{GB32}
H_0\sim H_l^6\equiv \frac{1}{192\beta\kappa^2}\ .
\ee
For the consistency $H_l\gg H_s$, hence $\alpha\beta\kappa^4\ll 1$.
Then the large curvature solution  (\ref{GB32}) might correspond to the inflation in
the early universe and the small curvature one in (\ref{GB31}) to the late time
acceleration. Similarly, one can construct many more explicit examples of
modified GB gravity which complies with solar system tests (at least, for
some backgrounds) and leads to
accelerating (effective cosmological constant, phantom or quintessence)
late-time cosmology.

\noindent
4. In summary, it is shown that modified GB gravity
 passes two fundamental
 solar system tests in the
same sense as General Relativity. It easily describes the late-time
acceleration of the universe whatever is the effective equation of state (
effective cosmological constant, quintessence or phantom). Moreover,
we demonstrated that such modified GB gravity may
 describe the transition
from deceleration to acceleration as well as crossing of phantom divide.
It remains to understand whether any limitations to
the function $f(G)$ from solar
system gravitational physics may be found. More serious limitations to its
form may be searched fitting the theory against the observational data.

\end{document}